\documentclass[conference]{IEEEtran}
\IEEEoverridecommandlockouts

\usepackage{cite}
\usepackage{amsmath,amssymb,amsfonts}
\usepackage{algorithmic}
\usepackage{graphicx}
\usepackage{textcomp}
\usepackage{xcolor}
\usepackage[ruled,vlined]{algorithm2e}
\usepackage{amsmath}
\usepackage{url}
\usepackage{array}
\usepackage{booktabs}
\usepackage{textcomp}
\usepackage{multirow} 
\usepackage{subcaption}
\usepackage{array}
\usepackage{enumitem}
\usepackage{fontawesome5}  
\usepackage{hyperref}

\def\BibTeX{{\rm B\kern-.05em{\sc i\kern-.025em b}\kern-.08em
    T\kern-.1667em\lower.7ex\hbox{E}\kern-.125emX}}
\begin{document}

\title{From Data Center IoT Telemetry to Data Analytics Chatbots – Virtual Knowledge Graph is All You Need}

\author{\IEEEauthorblockN{1\textsuperscript{st} Junaid Ahmed Khan}
\IEEEauthorblockA{\textit{DEI Department} \\
\textit{University of Bologna}\\
Bologna, Italy \\
junaidahmed.khan@unibo.it}
\and
\IEEEauthorblockN{2\textsuperscript{nd} Hiari Pizzini Cavagna}
\IEEEauthorblockA{\textit{DEI Department} \\
\textit{University of Bologna}\\
Bologna, Italy \\
hiari.pizzinicavagna@unibo.it}
\and
\IEEEauthorblockN{3\textsuperscript{rd} Andrea Proia}
\IEEEauthorblockA{\textit{DEI Department} \\
\textit{University of Bologna}\\
Bologna, Italy \\
andrea.proia2@unibo.it}
\and
\IEEEauthorblockN{4\textsuperscript{th} Andrea Bartolini}
\IEEEauthorblockA{\textit{DEI Department} \\
\textit{University of Bologna}\\
Bologna, Italy \\
a.bartolini@unibo.it}
}

\maketitle
    
\begin{abstract}

Industry~5.0 demands IoT systems that support seamless human–machine collaboration, yet current IoT data analysis requires deep domain, deployment, and query expertise. We show that combining Large Language Models (LLMs) with Knowledge Graphs (KGs) enables natural language access to heterogeneous IoT data. Focusing on data center IoT telemetry, we introduce a rule-based Virtual Knowledge Graph (VKG) construction process and an on-premise LLM inference service to create an end-to-end Data Analytics (DA) chatbot. Our system dynamically generates VKGs per query and translates user input into SPARQL, achieving 92.5\% accuracy (vs.\ 25\% for LLM-to-NoSQL) while reducing latency by 85\% (20.36\,s to 3.03\,s) and keeping VKG sizes under 179\,MiB. This work demonstrates that VKG-powered LLM interfaces deliver accurate, low-latency, and relationship-aware access to large-scale telemetry, bridging the gap between users and complex IoT systems in Industry~5.0.


\end{abstract}


\section{Introduction}

With the transition from the data‑driven Industry~4.0 to the personalized, collaborative, and sustainable Industry~5.0 production, the Internet of Things (IoT) must evolve by integrating people and processes with smart devices --- forming the Internet of Everything (IoE). Within this IoE ecosystem, a central objective is to reduce communication barriers between humans and smart devices, thus enabling seamless human‑machine collaboration and personalization \cite{SemIoE}. 
So far, we (humans) have learned to talk the language of the devices and in the context of IoT, we learned to write queries to NoSQL databases \cite{CACM-NoSQL} and/or to customize dashboards continuously displaying IoT data \cite{GRAFANA}. However, every user of an IoT system has struggled to extract value from the "stream" and "lake" of collected data, facing a three-fold barrier: (i) being a domain expert in the application field, (ii) being a domain expert in the IoT framework deployment and (iii) being an expert in the query language and API of the specific NoSQL database technology. 

What if we shift this paradigm and teach IoT systems to understand our natural language by leveraging recent progress in Large Language Models (LLMs)? What would be the key components that need to be added to a state-of-the-art IoT installation to enable this shift? And what would be the implications for cost and accuracy, in terms of computational resources and quality of communication? In this paper, we tackle these research questions in the context of data center telemetry, which applies IoT and Industry 4.0 technologies and principles to large-scale data center operations.

\sloppy

One of the fundamental challenges in realizing this vision is the heterogeneity of IoT data sources. Large-scale deployments integrate information from diverse devices, protocols, and data models, creating fragmentation that hinders interoperability. Knowledge Graph (KG) approaches have emerged as an effective solution, unifying these heterogeneous datasets within a shared semantic framework. By representing entities and their complex relationships as graphs, KGs enable more intuitive querying, interpretation, and integration of diverse data sources. This semantic unification proves especially valuable in data center telemetry, where KGs structure telemetry data into graph models that capture both physical and logical relationships, facilitating more efficient and intuitive data exploration through powerful graph query languages like SPARQL.

A recent paper by Zong et al. \cite{zong2024integratinglargelanguagemodels} evaluates the efficacy of Large Language Models (LLMs) in generating IoT sensor data analysis code on two relatively simple datasets, reporting that ChatGPT-4 achieves a success rate between 20\% and 50\% according to the natural language query complexity. More than 80\% of the errors are caused by the LLM failing in generating valid code for data fetching/import and using the right column name --- highlighting the limited capability of current SoA LLMs to ground themselves in IoT installation domain specifics.

In contrast, recent work \cite{khan2025exasage} has shown that KGs are effective for grounding LLMs in SPARQL query generation from natural language. The authors present a conversational interface to IoT data targeting data center telemetry by using an LLM to translate user input into SPARQL queries, leveraging the domain-specific ontology \cite{khanExaQueryProvingData2024}. Without a KG, the LLM achieves only 25\% accuracy—since this task requires explicitly handling the relationships between different NoSQL DB tables—whereas the KG improves this to 92.5\% by encoding these relationships directly within its graph structure. The authors also introduce a Virtual Knowledge Graph (VKG) approach for dynamic, query-driven KG construction, enabling scalable data center telemetry access. However, the reported 20.36s average end-to-end latency renders it impractical for real-time chatbot applications.

In this paper, we answer the above research questions by extending a SoA IoT framework targeting data center telemetry to turn it into an end-to-end Data Analytics (DA) Chatbot. For this purpose, we introduce (i) a rule-based VKG construction process, (ii) an on-premise LLM inference service, and (iii) an optimized KG creation pipeline. The chatbot architecture includes a frontend and backend: the frontend receives user questions and returns results in tabular and CSV formats; the backend, built with a Flask API, orchestrates VKG generation, LLM-based SPARQL query generation, and query refinement and execution on the graph database. We evaluate the chatbot’s effectiveness on the largest publicly available data center telemetry dataset \cite{m100nature}, where it achieves a response accuracy of 92.5\%. We provide a detailed characterization of the key components, the VKG generation process, and introduce optimization strategies for LLM inference and VKG generation, reducing the average response latency by 85\%—from 20.36s to 3.03s—compared to previous work \cite{khan2025exasage}. 
This paper builds on the emerging KG approaches to bridge the challenges of heterogeneous data representation and high-latency query access, enabling relationship-aware, intuitive, and low-latency access to data center IoT telemetry data through an integrated, graph-empowered natural language interface: from IoT dashboards to the first IoT chatbot. By releasing the source code, we aim to provide the community with a blueprint for extending this approach to other IoT installations.


The remainder of this paper is organized as follows. Section~\ref{sec:related} reviews the related works. Section~\ref{sec:chatbot} describes the proposed Data Analytics (DA) chatbot architecture and its components. Section~\ref{sec:frontend} presents the chatbot's frontend. Section~\ref{sec:backend} details the implementation of the backend and all its processes. Section~\ref{sec:results} presents the experimental results. Section~\ref{sec:current_limitations} discusses current limitations and future improvements. Finally, Section~\ref{sec:conclusion} concludes the paper. The DA chatbot is packaged and delivered as a docker container and is available at the Git repository: \url{https://gitlab.com/ecs-lab/da-chatbot}.

\section{Related Work}
\label{sec:related}

\subsection{Data Centers and Operational Data Analytics (ODA)}
\label{sec:oda_DCs}

Today’s data centers and high-performance computing (HPC) systems are no longer a niche but a driving industry sector concentrating large investments and building upon complex industrial plants. This sector is expected to reach nearly \$7 trillion in capital expenditures by 2030 worldwide to expand data center capacity and meet rising compute demands \cite{mckinsey2023costofcompute}. 
A data center today comprises hundreds of system racks and thousands of compute nodes built from heterogeneous multi-core CPUs, GPUs, and specialized accelerators. Equipped with millions of sensors and software metrics, they continuously generate heterogeneous telemetry streams \cite{borghesi2021examon} which are continuously being collected and stored during a data center's production aiming 
to improve efficiency and management \cite{netti2021conceptual}.

State-of-the-art data center telemetry frameworks, which are often referred to as Operational Data Analytics (ODA), rely on NoSQL-based databases \cite{ODA} to handle heterogeneous data sources while ensuring scalability and efficient storage. However, this flexibility comes with significant challenges in data querying due to the absence of a fixed schema, requiring users to manually establish connections between different data sources\cite{de2020sql,querying_heterogeneous_nosql,scherzinger2013managingschemaevolutionnosql}.  The large scale, component heterogeneity, and vendor-specific namespaces exacerbate the issue. For instance, the M100 dataset \cite{m100nature} (today the largest dataset of data centers IoT telemetry) from the Marconi-100 data center, deployed at the Cineca supercomputing facilities in Italy, captures telemetry from 49 system racks and 980 compute nodes. It integrates highly diverse data sources, ranging from hardware performance metrics (core loads, temperatures, frequencies, memory I/O, CPU power, fan speed, GPU usage) to facility infrastructure readings (liquid cooling, air conditioning, power supply units), workload statistics (job scheduling data, execution logs), operational alerts (system status events), and even external data such as weather forecasts. Collected over two and a half years, this mix of per-node metrics and system-wide measurements totals 49.9 TB of uncompressed data. Consequently, facility managers and data center operators struggle to acquire the knowledge needed to effectively query this data, limiting its potential \cite{khanExaQueryProvingData2024}. These challenges—data heterogeneity, lack of schema, and the difficulty of intuitive querying --- are not unique to data center telemetry but common to other IoT domains.

\subsection{Knowledge Graphs in IoT and Data Center Telemetry}

In IoT, the data is inherently heterogeneous, fragmented, and often lacks a shared schema. To address this, Knowledge Graphs (KGs) have emerged as a graph-based representation that integrates data from diverse sources into a unified semantic framework. KGs are typically expressed using the Resource Description Framework (RDF), which models information as subject–predicate–object triples. This allows entities (e.g., devices, users, sensors) and their relationships to be represented in a way that supports both human-readable and machine-interpretable querying through languages like SPARQL. The relationships and concepts within a KG are formally defined by an ontology, which acts as the schema by specifying the domain’s entities, their properties, and permissible connections.

KGs have been successfully applied across multiple IoT domains. For instance, in \cite{Iot_KG_useCase1} KGs are used to model group-level preferences in IoT recommendation systems, helping to represent complex interactions among diverse users, while in IoT security, KG-driven incremental learning techniques have shown promise in identifying specific devices or emitters with limited training data \cite{Iot_KG_useCase2}. Additionally, in \cite{iotKGMiddleware}, KGs are proposed as a semantic integration layer within IoT middleware to bridge communication gaps and unify heterogeneous device management. Furthermore, in \cite{knowledge_based_diagnosis_iiot}, ontologies and inference mechanisms are used to improve interpretability and fault diagnosis in IoT environments. Collectively, these works highlight the potential of KGs to unify heterogeneous data sources and facilitate interpretable insights critical for managing complex IoT ecosystems.

Recently, these graph-powered approaches have been applied to 
large-scale data center telemetry data. In \cite{khanExaQueryProvingData2024}, Khan et al. propose a KG-based approach for structuring data center telemetry data, accompanied by a domain-specific ontology to provide a well-defined schema to the telemetry data. The resultant KG represents monitored data in a graph structure, where vertices correspond to specific measurements and edges denote both topological and compositional relationships. This approach facilitates linking individual submitted jobs to the physical nodes on which they execute, as well as connecting nodes to the associated telemetry data. Authors of \cite{khanExaQueryProvingData2024} report that thanks to this KG representation, querying telemetry data is more intuitive for the end user, with SPARQL being the query language.

\subsection{Large Language Models for Query Generation}

While KGs enhance semantic organization and querying, accessing them through SPARQL still demands significant technical expertise. Large Language Models (LLMs) offer a natural language interface but struggle with factual accuracy and structured data access. LLMs excel in natural language understanding and code generation, but often produce hallucinated, non-factual outputs \cite{haluc_survey}. Integrating a KG into the query generation process has been shown to improve reliability by grounding LLMs in factual knowledge \cite{pan2023unifying,ji2021survey}. However, even state-of-the-art models like GPT-4 achieve only 54.89\% accuracy in real-world SQL query generation \cite{LLM-SQL-real-world}, and accuracy drops further when dealing with NoSQL databases for telemetry due to schema variability and data heterogeneity. A recent paper by Zong et al.\cite{zong2024integratinglargelanguagemodels} evalutates the efficacy of LLMs in generating IoT sensor data analysis code on two relative small datasets (tens of megabytes and few sensors’ types) with more than the 80\% of the errors being caused by the LLM failing in generating valid code for data fetching/import and using right column name. As stated earlier, this is a common problem of IoT systems also when used by humans, as they require three-fold domain expertise (knowing the installation-specific sensors’ name, their meaning, and the specific query language and data access API). 

Another emerging approach is Retrieval-Augmented Generation (RAG), where up-to-date textual information is stored as vector embeddings in a vector database. When a user submits a query, the system encodes it into a vector representation, retrieves the top-k most similar documents based on a similarity coefficient, and incorporates them into the LLM’s context to improve factual accuracy \cite{lewis2020retrieval}. While RAG enhances consistency by integrating external textual data, its reliance on text limits its ability to handle complex, non-textual (often timeseries) datasets common in IoT \cite{zong2024integratinglargelanguagemodels, m100nature}.


\subsection{The EXASAGE Framework}

In \cite{khan2025exasage}, the authors propose the EXASAGE framework, which extends the KG approach introduced in \cite{khanExaQueryProvingData2024} to enable a natural language LLM interface that translates natural language into SPARQL queries, achieving 92.5\% accuracy on data center telemetry data. This high accuracy is obtained by providing the KG schema and a few-shot examples in the LLM context and parsing the generated SPARQL query through a query refinement script which removes common SPARQL syntax errors. The EXASAGE framework reports an average end-to-end latency of 20.36s per prompt, with VKG generation taking 8.05s (approximately 39.5\% of total latency) and LLM inference accounting for 11.09s (approximately 45.5\%).

While EXASAGE obtains high query accuracy, it does not provide a chatbot implementation. Indeed, this high latency limits its practicality for real-time interaction with large-scale data, common in data center telemetry. In contrast, this paper proposes a complete end-to-end implementation of a chatbot, with system-level optimizations to reduce latency through improvements in VKG generation and LLM inference efficiency. These enhancements make our solution more viable for interactive, graph-based querying of telemetry data in real-world IoT environments.

\section{Data Analytics (DA) Chatbot Architecture}
\label{sec:chatbot}

\begin{figure*}[ht]
    \centering
    \includegraphics[width=\linewidth]{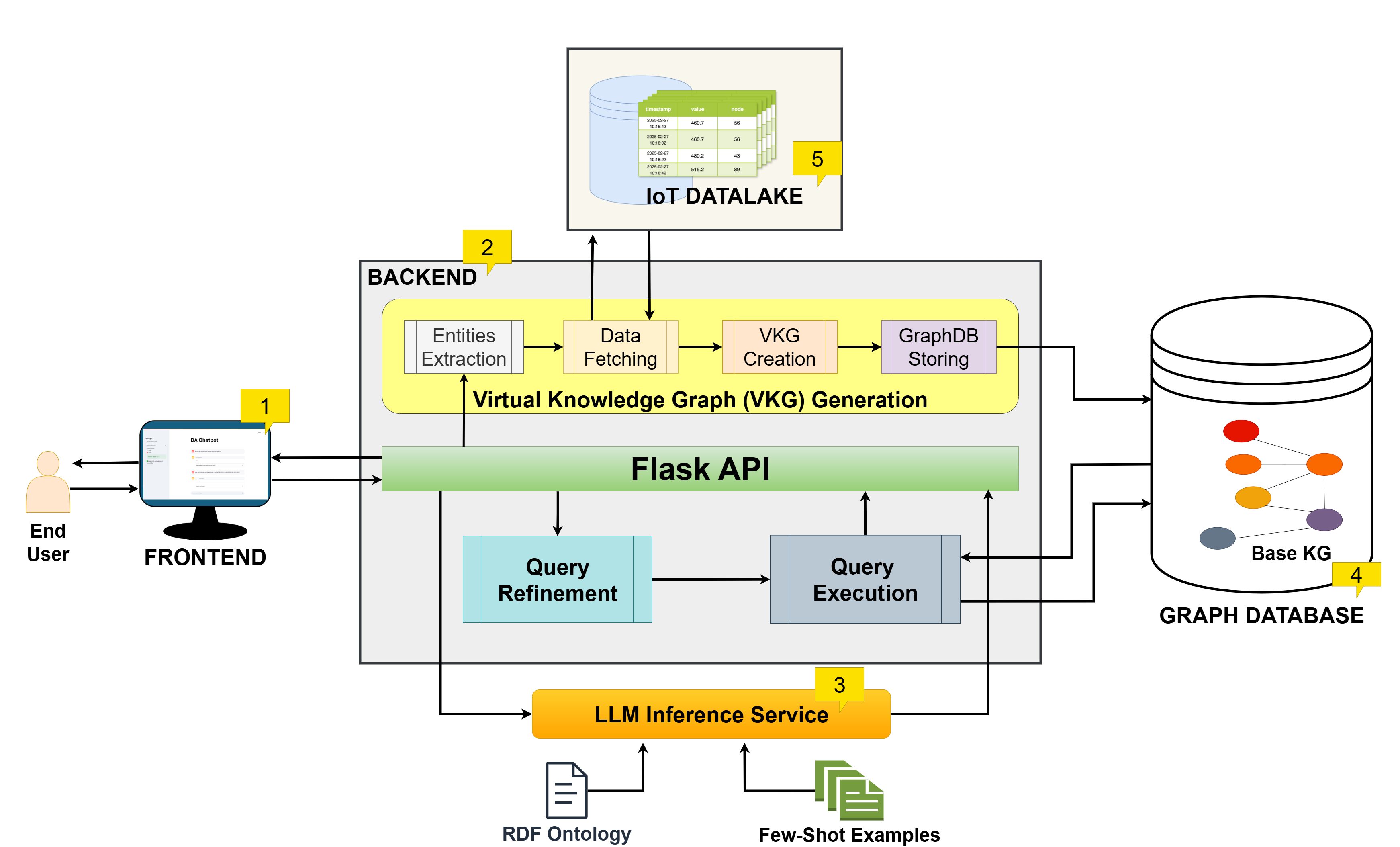}
    \caption{Data Analytics (DA) Chatbot Architecture Block Diagram.}
    \label{fig:oda_chatbot}
\end{figure*}

Figure~\ref{fig:oda_chatbot} shows the proposed Data Analytics (DA) Chatbot architecture, which consists of five components:
\begin{enumerate}[left=0pt, label=\textbf{\arabic*.}, labelsep=0.5em, itemsep=0.5em, topsep=0.5em, partopsep=0.5em]
    \item \textbf{Frontend:} The frontend is a chat-based user interface built using the Python framework Streamlit. It enables users to interact with telemetry data by submitting queries in natural language. These queries are forwarded to the backend for processing, and the results are returned to the frontend, where they are displayed either as a table or as a downloadable CSV file.
    \item \textbf{Backend:} Built with the Flask API, the backend serves as the central orchestrator of the chatbot architecture. Upon receiving a user's question from the frontend, it initiates two parallel processes: (a) LLM Inference – the prompt is sent to the LLM inference service, which generates a SPARQL query based on the KG schema and a few-shot context; and (b) VKG Generation – a temporary KG is constructed, containing only the data necessary to answer the user’s query. 
    Once both tasks are completed, the backend proceeds to the query refinement step—similar to \cite{khan2025exasage}—where the SPARQL query is validated and corrected for common syntax errors or invalid property references. The refined query is then executed against the Graph Database’s SPARQL endpoint. The final result is returned to the frontend for presentation to the user.
    \item \textbf{LLM Inference Service:} This component is responsible for generating SPARQL queries for user questions. It constructs the LLM's input by combining the user text with two key contextual elements: (i) a textual representation of the underlying KG schema of the telemetry in the form of an RDF ontology. This ontology provides the schema-level knowledge upon which the VKG is constructed in the backend, and (ii) a set of few-shot examples demonstrating correct query generation. The (i) and (ii) guide the LLM to generate queries that are both syntactically accurate and semantically consistent with the KG schema.
    \item \textbf{Graph Database:} The graph database serves as the storage backend for dynamically generated IoT telemetry VKGs. It also stores a static KG, termed the \textit{Base-KG}, which is built according to the domain-specific RDF ontology. This \textit{Base-KG} contains domain-specific system-level configurations and metadata—such as static IoT sensor configurations, device hierarchies, and spatial arrangements—that are specific to each deployment and must be adjusted accordingly. By combining the \textit{Base-KG} with query-specific VKGs, the system fulfills all the requirements for successful execution of the generated SPARQL query. This component can be implemented using any high-performance graph store, such as GraphDB by Ontotext, Apache Jena TDB/Fuseki, etc.
    \item  \textbf{IoT Datalake:} This is a NoSQL DB that contains the IoT telemetry data. The backend utilizes this IoT datalake for the construction of dynamic VKGs. It can either represent the production database \cite{ODA} or a repository of historical IoT datasets. 
\end{enumerate}

\section{Frontend}
\label{sec:frontend}



\begin{figure*}[htbp]
  \centering
  \begin{subcaptionbox}{Setup page\label{fig:setup_page}}[0.44\textwidth]
    {\fbox{\includegraphics[width=\linewidth, height=4.6cm]{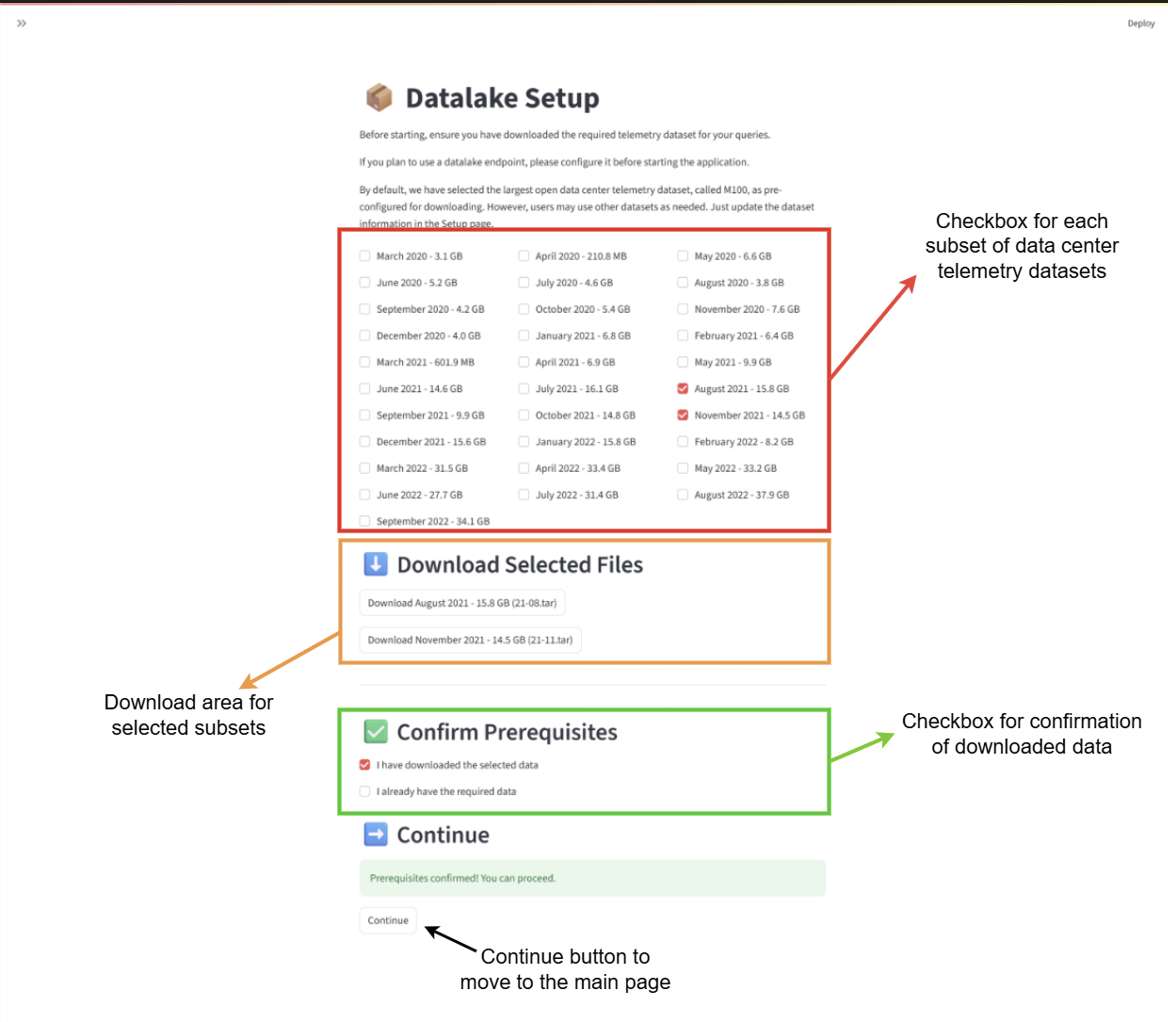}}}
  \end{subcaptionbox}
  \hfill
  \begin{subcaptionbox}{Main page\label{fig:main_page}}[0.44\textwidth]
    {\fbox{\includegraphics[width=\linewidth]{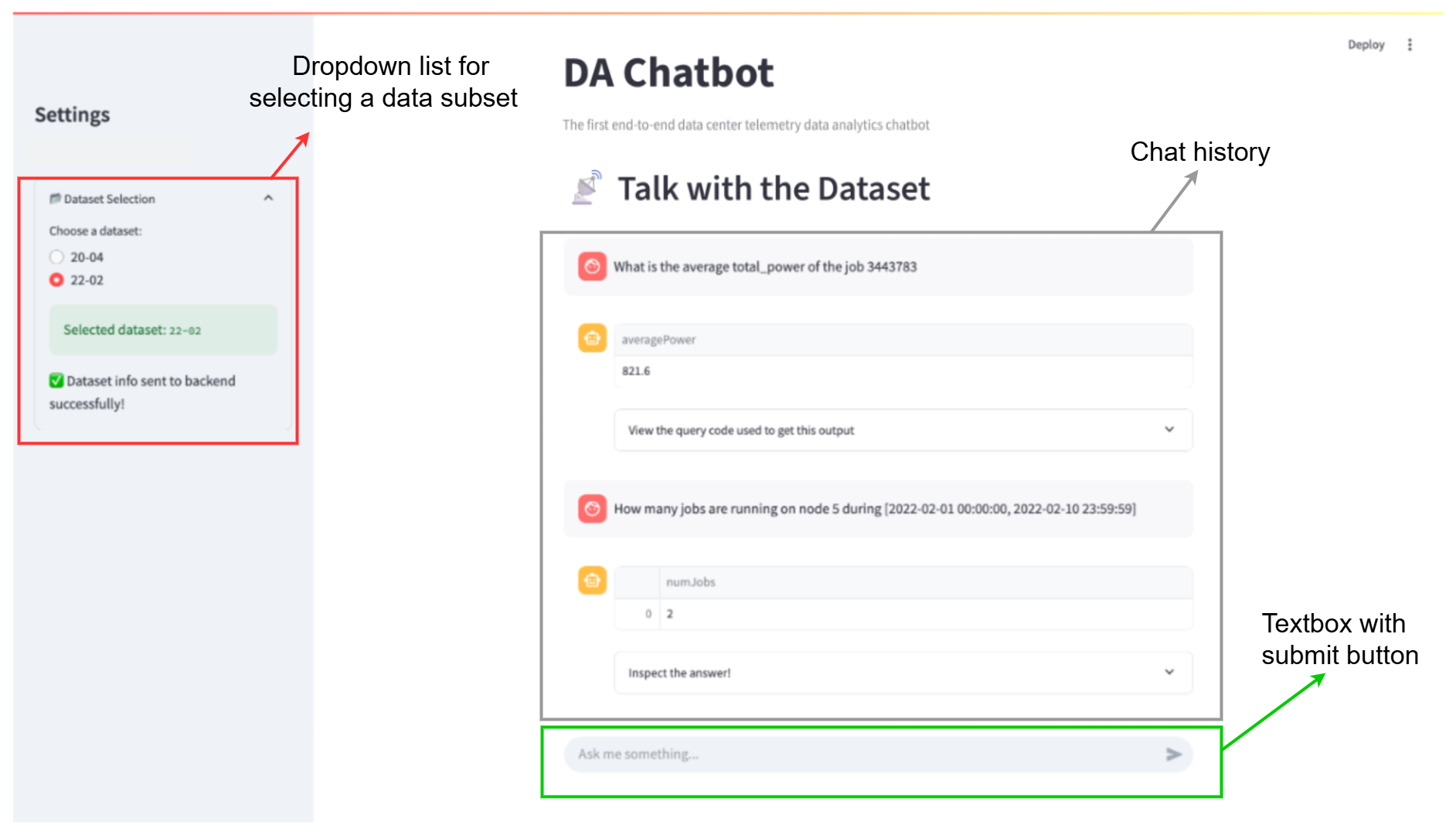}}}
  \end{subcaptionbox}
  \caption{Frontend built using Python framework Streamlit containing a setup page and a main page.}
  \label{fig:frontend}
\end{figure*}
The proposed frontend has been designed to facilitate end-user access to IoT data. As shown in Figure~\ref{fig:frontend}, it comprises two pages: (i) a setup page for configuring the IoT datalake and, in the case of historical traces, downloading them locally from their data repository, and (ii) a main page featuring a chatbot interface through which users can submit their questions for data analytics.

\subsection{Setup page}

Long-term storage of large-scale IoT datasets often requires sharding and chunking due to the substantial size and complexity of the data. Such datasets are typically divided into multiple subsets to facilitate manageable access and download \cite{m100nature,belarbi2025gothamdataset2025reproducible,mbc1_1h68_22}.

This setup page simplifies the process of locating and retrieving these subsets by presenting three main sections: (i) a checklist displaying all available subsets of the selected IoT dataset, along with relevant metadata such as size and time span; (ii) a download area where, upon selection, a download button appears for each chosen subset, allowing users to initiate data retrieval requests from the underlying data repository; and (iii) a confirmation section where users verify that the required data has been downloaded, either by confirming “I have downloaded the selected data” or “I already have the required data” if previously obtained. Users who intend to interact with the IoT datalake endpoint directly may bypass the download process by selecting the latter confirmation option; however, this requires prior configuration of the IoT datalake connection parameters. Once confirmation is provided, users can proceed to the main page by clicking the “Continue” button.

\subsection{Main page}

Figure~\ref{fig:main_page} illustrates the main page containing the chat interface of the chatbot, which is organized into three sections: (i) A dropdown list for selecting the IoT datalake endpoint or data subset among the ones available locally. Once selected, the dataset information is sent to the backend and processed by the Flask API's dataset route; (ii) A textbox with a submit button that allows users to enter and submit their data center telemetry questions. When clicked, the submit button sends the query to the Flask API's chat route for backend processing; and (iii) A chat history section that listens to the chat route of the Flask API and displays the results of submitted questions as a Pandas DataFrame. If the result contains more than 50 rows, only the first five are shown, with an option to download the full results as a CSV file. This section enables users to track their previous queries and corresponding responses. The implementation details of both the dataset and chat routes are described in the next section.

\section{Backend}
\label{sec:backend}

The backend of the DA Chatbot is implemented as a Flask API, serving as the central orchestration layer for handling data center telemetry-related user questions. It exposes two primary POST endpoints:
\begin{itemize}
    \item \textbf{dataset route:} This endpoint accepts the IoT datalake configuration and data subset as specified by the user. It provides context to the VKG generation by selecting the appropriate data source for retrieval.
    \item  \textbf{chat route:} This endpoint handles the submitted user questions in natural language. Upon receiving a request, the backend concurrently initiates two processes:
        \begin{itemize}
            \item \textit{VKG generation:} Utilizes the dataset parameters from the dataset route and the user's question to retrieve and structure the appropriate subset of data from the IoT datalake into a KG for storing in the graph database.
            \item \textit{LLM inference:} Utilizes the LLM inference service of the DA chatbot system to generate the SPARQL query based on the user's question.
        \end{itemize}
        Once both the \textit{VKG generation} and \textit{LLM inference} processes are completed, the backend forwards the SPARQL query generated by the \textit{LLM inference} to the \textit{Query refinement} process. During this stage, the generated query is validated and corrected for common syntactic and semantic errors to ensure its successful execution. Following this in the \textit{Query execution} process, the refined SPARQL query is then submitted to the graph database via its SPARQL endpoint. Upon successful execution, the query results are returned to the backend, which subsequently transmits the answer to the frontend through the chat route for presentation to the end user.
\end{itemize}

In the following, we detail the implementation of the three processes: \textit{VKG generation}, \textit{Query refinement}, and \textit{Query execution}. While \textit{LLM inference} is a key step in generating the SPARQL query, its implementation focuses on optimizing model execution through deployment configurations. Details of these configurations can be found in Section \ref{sub_sec:experimental_setting}. While the proposed architecture is applicable to any IoT installation, in this paper, we present an implementation that is specific to data center IoT telemetry and ODA. The following sections will use this domain-specific configuration.

\subsection{Virtual Knowledge Graph (VKG) Generation}
\label{sec:VKG}

The proposed VKG process consists of four sub-tasks: (1) Entity Extraction, (2) Data Fetching, (3) VKG Creation, and (4) GraphDB Storing. In the DA Chatbot, the VKG is dynamically constructed using only data relevant to a specific user query expressed in natural language (hereafter referred to as the user’s question). The process begins with interpreting the user’s intent by identifying the entities referenced in the question, which guide subsequent data retrieval and graph construction. Algorithm~\ref{alg:process-prompt} outlines the pseudocode for the Entity Extraction task, which identifies relevant entities from the user’s question. These entities are extracted based on predefined categories aligned with the domain-specific data center telemetry ontology \cite{khan2025exasage}. The categories include ontology classes—node, rack, job, metric, and plugin—as well as temporal markers: start time and end time, to support time-based aggregations or intervals. A key characteristic of telemetry data is its timeseries nature, meaning user questions may include (i) explicit time intervals or (ii) implicit ones corresponding to job execution periods. The algorithm begins by converting the user’s question to lowercase and initializing an entities dictionary. Each key corresponds to a predefined category and is initialized with a “present” flag set to False and a “value” set to None. The “present” key indicates whether the entity is mentioned in the question, while “value” stores any associated content. The algorithm iterates over the categories, checking for their presence in the user question. If found, it marks the entity as present and extracts its value using Python regular expression matching. If node mappings are available, a node mapping function can be used to standardize node identifiers according to facility-specific conventions. The final output is a Python dictionary containing the extracted entities, referred to as entities.

\begin{algorithm}[h]
\footnotesize
\caption{Entity Extraction from $user\_question$}
\label{alg:process-prompt}
\KwIn{$user\_question$, $plugin\_metric\_map$, optional $node\_mappings$}
\KwOut{$entities$ dictionary}

Convert $user\_question$ to lowercase\;  
Define categories: \{\texttt{node, rack, job, metric, plugin, start\_time, end\_time}\}\;  
Initialize $entities[cat]$: \texttt{present} $\gets$ \texttt{False}, \texttt{value} $\gets$ \texttt{None}\;

\ForEach{$cat$ in categories}{
    \If{$cat$ in $user\_question$}{
        $entities[cat][present] \gets$ \texttt{True}\;  
        Extract value via regex $\rightarrow$ $val$\;
        \If{$val$ found}{$entities[cat][value] \gets val$\;}
        \If{$cat = $ \texttt{node} and $node\_mappings$ exists}{Map using $node\_mappings$\;}
        \If{$cat = $ \texttt{metric}}{
            Identify plugin via $plugin\_metric\_map$\;  
            $entities[plugin][present] \gets$ \texttt{True}; $entities[plugin][value] \gets$ plugin\;
        }
    }
}
\Return $entities$\;
\end{algorithm}

Once these entities are identified, the next task is to fetch the relevant data from the IoT datalake and construct the VKG. The IoT datalake contains the telemetry data of the target data center. Algorithm \ref{alg:vkg-generator} presents the pseudocode for Data Fetching task and VKG Creation task. The VKG is constructed using RDF, and it adheres to the schema defined by the data center telemetry ontology. The algorithm begins by identifying key entities from the output of Algorithm~\ref{alg:process-prompt}. Key entities are those whose “present” flag in the entities dictionary is set to True. If no key entities are found, the algorithm raises an error. If only rack or node entities are present, VKG generation is skipped. Such topological user questions can be answered directly using the \textit{Base-KG}. If a job entity is present, the algorithm retrieves job data using either a job Id or a time range (start and end time), based on the information available in the value field of the entities dictionary. It then generates RDF triples representing the job and establishes links to its associated compute nodes, applying node mappings if available. For metric entities, the algorithm first ensures that a valid time range is provided before retrieving sensor readings. It then fetches readings either for a specified node or for all nodes if none is indicated, generates RDF triples for the reading values, and associates the readings with the relevant plugins and sensors. The constructed VKG is then returned as the output of the algorithm.


\begin{algorithm}[h]
\footnotesize
\caption{Virtual Knowledge Graph (VKG) Generation}
\label{alg:vkg-generator}
\KwIn{$entities$, optional $node\_mappings$}
\KwOut{$VKG$}

$key\_entities \gets \{ e \mid entities[e][present] = \texttt{True} \}$\;

\If{$key\_entities = \emptyset$}{
    \textbf{Raise} \texttt{ValueError} (``No valid key\_entities found'')\;
}
\If{$key\_entities \subseteq \{\texttt{rack, node}\}$}{
    \Return\;
}
\If{\texttt{job} $\in$ $key\_entities$}{
    \If{$entities[job][value] \neq$ \texttt{None}}{
        $jobId \gets entities[job][value]$\;
        Retrieve job data for $jobId$ from IoT datalake\;
    }
    \ElseIf{\texttt{start\_time}, \texttt{end\_time} $\in key\_entities$}{
        $startTime \gets entities[start\_time][value]$\;
        $endTime \gets entities[end\_time][value]$\;
        Retrieve job data from IoT datalake between $startTime$ and $endTime$\;
    }
    \Else{
        \textbf{Raise} \texttt{ValueError} (``Provide jobId or time range'')\;
    }
    Add RDF triples for job attributes ($jobId$, $startTime$, $endTime$) in $VKG$\;
    \If{$node\_mappings$ exists}{Apply to node names\;}
    Add job–node relations in $VKG$\;
}
\If{\texttt{metric} $\in$ $key\_entities$}{
    \If{\texttt{start\_time} $\notin key\_entities$ or \texttt{end\_time} $\notin key\_entities$}{
        \If{\texttt{job} $\in key\_entities$}{Use job's time range\;}
        \Else{\textbf{Raise} \texttt{ValueError} (``Provide $start\_time$ and $end\_time$'')\;}
    }
    $metric \gets entities[metric][value]$,\;
    $plugin \gets entities[plugin][value]$\;

    $startTime \gets entities[start\_time][value]$\;
    $endTime \gets entities[end\_time][value]$\;
    
    \If{$entities[node][value] \neq$ \texttt{None}}{
        $node \gets entities[node][value]$\;
        Retrieve metric data for $node$ from IoT datalake between $startTime$–$endTime$\;
    }
    \Else{
        Retrieve all metric data between $startTime$–$endTime$ from IoT datalake\;
    }
    
    \If{$node\_mappings$ exists}{Apply to node names\;}

    \ForEach{$node$}{Add RDF triples in $VKG$ for node, plugin, sensor\;}
    \ForEach{$reading$}{Add RDF triples: ($value$, $timestamp$, $unit$) linked to sensor\;}
}
\Return $VKG$\;
\end{algorithm}

Upon successful execution of Algorithm~\ref{alg:vkg-generator}, both the Data Fetching and VKG Creation tasks are completed. The resulting VKG contains all the necessary data to answer the user's question. The final task is GraphDB Storing, in which the VKG is persisted in the graph database. This ensures that the data is readily accessible for execution of the SPARQL query generated by the LLM inference service.

\subsection{Query Refinement (QR)}
\label{sec:query_refinement}

Since LLMs are known to be prone to hallucinations \cite{haluc_survey}, a validation process was necessary to identify and correct erroneously generated SPARQL queries. To address this, we adopted the query refinement process from its implementation in the EXASAGE framework \cite{khan2025exasage}, where authors employed Python regular expression (regex) matching to detect and correct common categories of errors, including improper data type usage and the introduction of non-existent entities or relations not defined in the data center telemetry ontology. This refinement process ensures that the generated SPARQL queries are syntactically valid and semantically aligned with the underlying KG schema.

\subsection{Query Execution (QE)}
\label{sec:query_execution}

The query execution takes as input the refined query produced by the query refinement process and submits it to the SPARQL endpoint of the graph database for execution. It then waits for a response from the SPARQL endpoint, which contains the answer to the user's question. Once the response is received, it is sent to the Flask API, which subsequently forwards the answer to the frontend using the chat route to display it to the end user.




\section{Experimental Results}
\label{sec:results}




In this section, we evaluate the proposed DA Chatbot by exploring different implementation configurations and to answer the research questions. We then apply a series of optimizations to both the VKG generation and LLM inference to improve the average response time of the system (also referred to as end-to-end latency throughout the paper).

\subsection{Experimental setting}
\label{sub_sec:experimental_setting}

All experimental results were obtained using three distinct computing resources: 
(1) For the IoT Datalake, we consider two configurations: (i) Production NoSQL DB – The Examon holistic data center telemetry monitoring framework that was deployed at CINECA during the operational phase of the Marconi100 (M100) supercomputer. It ran on the CINECA Ada Cloud, using Cassandra as the NoSQL DB and KairosDB as its time-series extension, (ii) Parquet files – We use publicly available historical data center telemetry data for the M100 supercomputer from Zenodo \cite{m100nature}, specifically the Parquet files from February 2022.
(2) For the Graph database, we selected the GraphDB-Free edition by Ontotext 
and the \textit{Base-KG} stores metadata about the M100 data center—such as spatial layout, rack configurations, and compute node locations; and 
(3) The DA Chatbot experiments were conducted on a server equipped with an Intel Xeon Platinum 8480+ CPU, 2.0 TiB of RAM, and an NVIDIA H100 80 GB HBM3 GPU, using TensorRT-LLM v0.19\footnote{\href{https://nvidia.github.io/TensorRT-LLM/overview.html}{TensorRT-LLM documentation}}, an NVIDIA framework designed to optimize and deploy LLMs leveraging several optimizations. Specifically, we executed TensorRT-LLM within a Docker container and compiled the LLaMA 3.1 8B Instruct model with both tensor parallelism and pipeline parallelism set to 1 to use a single GPU. 

\subsection{User questions}
\label{sec:user_prompts}

Table \ref{tab:prompts} presents the twelve query archetypes used to evaluate the proposed DA Chatbot. These archetypes were used as templates to generate hundred user questions by replacing placeholder values with appropriate values. For each question, start and end times were randomly selected to span one month. 

\begin{table}[ht]
\centering
\caption{Domain-specific data center telemetry complex query archetypes for evaluating the DA Chatbot with placeholders for dynamic query parameters.}
 \label{tab:prompts}
\begin{tabular}{|c|p{7.5cm}|}
\hline
\textbf{Id} & \textbf{User Question} \\ \hline
1 & Which nodes are present in the rack [rack\_id], and what are their positions? \\ \hline
2 & What were the nodes used by the job [job\_id]? \\ \hline
3 & What is the average [metric] consumption for the job [job\_id]? \\ \hline
4 & Which jobs had execution time higher than [minutes] minutes and were submitted between [start\_time] and [end\_time]? \\ \hline
5 & How many jobs were running on the node [node\_id] between [start\_time] and [end\_time]? \\ \hline
6 & What is the maximum, minimum, and average [metric] of the node [node] between [start\_time] and [end\_time]? \\ \hline
7 & How many jobs were running on the rack [rack\_id] between [start\_time] and [end\_time]? \\ \hline
8 & Identify the nodes that exceeded the [metric] consumption threshold of [threshold] between [start\_time] and [end\_time]. \\ \hline
9 & What is the average execution time of the jobs submitted between [start\_time] and [end\_time]? \\ \hline
10 & List all jobs running between [start\_time] and [end\_time]. \\ \hline
11 & What is the average [metric] of node [node\_id] between [start\_time] and [end\_time]? \\ \hline
12 & What is the duration of jobs between [start\_time] and [end\_time]? \\ \hline
\end{tabular}
\end{table}

\subsection{Data Analytics (DA) Chatbot Implementation}
\label{sec:oda_chatbot_implementations}

\begin{table*}[ht]
\centering
\caption{Comparative evaluation of DA Chatbot deployment configurations, reporting the average end-to-end latency, answer accuracy, and storage size overhead across different IoT datalake configurations (NoSQL DB and Parquet files). The end-to-end latency includes the measured LLM inference latency (as in Table~\ref{tab:llm_inference_metric}), and the query refinement and query execution latencies (as in Table~\ref{tab:qr+qe}).}
\label{tab:chatbot_implementations}
\resizebox{\textwidth}{!}{%
\begin{tabular}{|l|r|r|r|r|}
\hline
\textbf{DA Chatbot Configuration} & \textbf{IoT Datalake} & \textbf{End-to-End Latency [s]} & \textbf{Accuracy [\%]} & \textbf{Storage Size [GiB]} \\ \hline
LLM-to-NoSQL (DA Chatbot + \cite{ODA}) & NoSQL DB & 102.54 & 25.0 & - \\ \hline 
LLM-to-SPARQL (DA Chatbot + \cite{khanExaQueryProvingData2024}) & - & 1.42 & 92.5 & 2979.84 \\ \hline 
LLM-to-SPARQL via VKG-Naive (DA Chatbot + \cite{khan2025exasage}) & NoSQL DB & 9.04 & 92.5 & 0.17 \\ \hline 
LLM-to-SPARQL via VKG-Naive (DA Chatbot + \cite{khan2025exasage}) & Parquet & 6.92 & 92.5 &  0.17 \\ \hline 
\end{tabular}%
}
\end{table*}


Given the two IoT datalake formats—NoSQL DB for the production data center telemetry DB and Parquet files for historical data center telemetry—the DA Chatbot can be implemented in four distinct configurations: 
(1) \textbf{LLM-to-NoSQL Query Generation}: The LLM directly generates NoSQL query code, executed against a NoSQL DB used as the IoT datalake. This represents a naive implementation of a DA chatbot leveraging state-of-the-art (SoA) data center telemetry technologies \cite{ODA} and the LLM’s internal knowledge. Comparisons against this highlight the benefits of KG representation;
(2) \textbf{LLM-to-SPARQL on Data Center Telemetry KG}: The LLM generates SPARQL queries executed on a pre-built data center telemetry KG. In this configuration, there is no requirement for an IoT datalake to construct a VKG, as a full KG has already been built containing all the data center IoT telemetry data. This setup represents a DA chatbot based on the data center telemetry ontology \cite{khanExaQueryProvingData2024} as the schema for the pre-built KG. It represents the lower bound of performance but is unfeasible due to KG storage requirements;
(3) \textbf{LLM-to-SPARQL via VKG (NoSQL DB)}: A VKG generator with a NoSQL DB as the IoT datalake, allowing the LLM to generate SPARQL queries that are resolved through this VKG layer;
(4) \textbf{LLM-to-SPARQL via VKG (Parquet)}: Similar to configuration (iii), but the VKG is constructed using Parquet files as the IoT datalake. 

As the VKG, we used the implementation proposed in \cite{khan2025exasage}, which is available at (https://gitlab.com/ecs-lab/exasage), but with the optimized LLM inference proposed in this paper. To identify the most optimal deployment configuration, we conducted a comparative evaluation using randomly generated user questions (see Section~\ref{sec:user_prompts}), assessing performance based on average values across the following metrics: 
(i) End-to-End Latency, measured as the average response time from user question to final answer, including the cumulative duration of all processes in the DA Chatbot: LLM inference, VKG generation (in configurations (iii) and (iv)), query refinement, and query execution;
(ii) Answer accuracy – evaluating the correctness and completeness of the DA Chatbot’s response to the user's questions; and
(iii) Storage size overhead – storage requirements for each deployment configuration.

The results of this evaluation are summarized in Table~\ref{tab:chatbot_implementations}. The LLM-to-NoSQL approach, while straightforward, suffers from a high end-to-end latency of 102.54s and a very low accuracy of just 25\% in correctly answering user questions. Of the 102.54s end-to-end latency, only 2.9\% is related to LLM inference. The high latency is primarily due to the complexity of answering user questions that involve accessing multiple heterogeneous data sources. This is exacerbated when questions require multiple sub-queries to assemble the final answer. For example, to calculate the average power consumption of all jobs running within a time period \textit{T}, the query engine must query the job table for relevant jobs and retrieve corresponding data from power consumption sensor tables for each job’s time span, leading to significant query execution overhead in the DA Chatbot. The LLM-to-NoSQL results demonstrate that (i) latency is dominated by the data center telemetry DB, (ii) the accuracy makes this naive implementation useless. Recent work \cite{LLM-SQL-real-world} shows that even using an SQL-capable data center telemetry DB, accuracy would be at a maximum of 54.9\% -- This motivates the proposed VKG approach as an essential innovation for DA Chatbots. In terms of storage, it does not require any additional storage beyond the one already used for data center telemetry. 

Indeed, in the second approach, LLM-to-SPARQL (with a pre-built data center telemetry KG, constructed using the IPMI plugin data from February 2022 \cite{m100nature}), the end-to-end latency drops drastically to 1.42s, and the accuracy improves significantly to 92.5\%—a 67.5\% gain in accuracy and a 98.6\% reduction in latency compared to the LLM-to-NoSQL configuration. While this configuration offers excellent performance, it is impractical in terms of storage. The complete data center telemetry KG for just the IPMI plugin data from one month occupies 2979.84 GiB—approximately 745× larger than its Parquet-based NoSQL equivalent, which occupies only 4.00 GiB. 

The two above approaches make clear that DA chatbots are impossible with SoA technologies --- (i) today's LLMs cannot generate accurate data center telemetry queries from natural language (25\% of accuracy) and (ii) KG representation of data center telemetry data are effective in grounding LLMs (92.5\% of accuracy) but unpractical in terms of storage (745x than parquet files) --- motivating the proposed VKG approach (LLM-to-SPARQL via VKG), where a KG is built on-demand using only the data necessary to answer each user question. For this evaluation, we adopted the VKG implementation in \cite{khan2025exasage}, and we will henceforth term it as VKG-Naive. This approach retains the advantages of using a KG—such as high accuracy and low latency—while significantly reducing storage overhead. We evaluate this method under both IoT datalake configurations. In the NoSQL DB IoT datalake configuration, the VKG approach achieves an end-to-end latency of 9.04s, with accuracy remaining high at 92.5\%. Although the additional VKG construction step increases latency by 84.3\% compared to the pre-built KG configuration, it makes the implementation practically feasible: the maximum storage overhead across all user questions is just 0.17 GiB, which is negligible compared to the 2979.84 GiB required by the complete data center telemetry KG. Furthermore, in the Parquet IoT datalake configuration, the end-to-end latency gets to 6.92s—a 23.5\% improvement over its NoSQL DB IoT datalake counterpart, and a 93.3\% improvement over the original LLM-to-NoSQL setup. Importantly, the accuracy (92.5\%) and maximum storage overhead (0.17 GiB) remain unchanged.

\subsection{Benchmarking LLM Inference Service performance in the DA Chatbot}
\label{sec:llm_inference}

To assess the performance of the LLM Inference Service within the DA Chatbot, we benchmarked the LLaMA 3.1 8B Instruct model using TensorRT-LLM. Table \ref{tab:llm_inference_metric} summarizes the LLM inference metrics of Total Output Throughput (tokens/s) and the Total LLM Inference Latency (s) for the two configurations of LLM-to-NoSQL and LLM-to-SPARQL.

\begin{table}[ht]
\centering
\caption{Average output tokens per second and average LLaMA 3.1 8B Instruct inference latency (with HW setup as in Section~\ref{sub_sec:experimental_setting}) for both LLM-to-NoSQL and LLM-to-SPARQL configurations.}
\label{tab:llm_inference_metric}
\resizebox{\columnwidth}{!}{%
\begin{tabular}{|l|r|r|}
\hline
\textbf{Metric}                        & \textbf{LLM-to-NoSQL} & \textbf{LLM-to-SPARQL} \\ \hline
Input Tokens                           & 2935           & 1952            \\ \hline
Output Tokens                          & 350            & 151             \\ \hline
Total Output Throughput {[}tokens/s{]} & 117.58         & 117.99          \\ \hline
Total LLM Inference Latency {[}s{]}    & 2.98           & 1.28            \\ \hline
\end{tabular}%
}
\end{table}

The average input and output token counts for the LLM-to-NoSQL configuration were 2,935 and 350, respectively. Under this setup, the total output throughput was 117.58 tokens/s, and the average LLM inference latency was measured at 2.98s. In contrast, for the second configuration, LLM-to-SPARQL, the average input and output tokens were 1,952 and 151, respectively. This illustrates a key benefit of the KG-based approach: SPARQL query code is significantly more concise than NoSQL, with an average reduction in length by 56.89\%. While the total output throughput remains nearly the same, as expected, the conciseness of SPARQL results in a reduced average LLM inference latency of 1.28s.


Compared to \cite{khan2025exasage}, where the LLM inference times were 11.09s for SPARQL and 16.09s for NoSQL, the proposed configuration achieves a substantial reduction in inference time—by 88\% for LLM-to-SPARQL and 81\% for LLM-to-NoSQL.

\subsection{Benchmarking Query Refinement (QR) and Query Execution (QE) in the DA Chatbot}
\label{sec:qr+qe}

\begin{table}[ht]
\centering
\caption{Average latencies for QR and QE for both LLM-to-NoSQL and LLM-to-SPARQL configurations.}
\label{tab:qr+qe}
\resizebox{0.6\columnwidth}{!}{%
\begin{tabular}{|l|r|r|}
\hline
\textbf{Configuration} & \textbf{QR{[}ms{]}} & \textbf{QE   {[}s{]}} \\ \hline
LLM-to-NoSQL           & -                   & 99.56                 \\ \hline
LLM-to-SPARQL          & $<0.01$                & 0.14                  \\ \hline
\end{tabular}%
}
\end{table}

Table~\ref{tab:qr+qe} presents the average latencies for query refinement and query execution across the two configurations. In the LLM-to-NoSQL setup, no refinement step is applied, and the query is executed directly, resulting in a high QE latency of 99.56s (due to the reasons identified in section~\ref{sec:oda_chatbot_implementations}). In contrast, LLM-to-SPARQL includes a minimal QR step ($<0.01$ ms) and achieves a significantly lower QE latency of 0.14s, highlighting the efficiency of the SPARQL and graph-based approach.

\subsection{Virtual Knowledge Graph (VKG) generation optimizations}
\label{sec:perf_vkg}

\begin{table*}[ht]
\centering
\caption{Latency breakdown by sub-tasks for the naive and optimized VKG implementations on the two types of IoT datalakes (NoSQL DB/Parquet), where the VKG-Naive is the implementation used in Table~\ref{tab:chatbot_implementations}.}
\label{tab:latency_vkg_versions}
\resizebox{\textwidth}{!}{%
\begin{tabular}{|l|r|r|r|r|r|r|}
\hline
\textbf{VKG   Version} & \textbf{IoT Datalake} & \textbf{Entities   Extraction [s]} & \textbf{Data Fetching   [s]} & \textbf{VKG Creation [s]} & \textbf{GraphDB Storing   [s]} & \textbf{Total VKG Latency [s]} \\ \hline
VKG-Naive & NoSQL DB & 0.00001 & 3.72559 & 3.92415 & 1.25232 & 8.90207 \\ \hline
VKG-Naive & Parquet & 0.00001 & 1.66915 & 3.86718 & 1.24132 & 6.77766 \\ \hline
VKG-Optimized & NoSQL DB & 0.00001 & 3.73122 & 1.53934 & 1.24995 & 6.52052 \\ \hline
VKG-Optimized & Parquet & 0.00001 & 0.12104 & 1.53804 & 1.24045 & 2.89954 \\ \hline
\end{tabular}%
}
\end{table*}

In this subsection, we evaluate the optimizations we propose in this paper to overcome the limitations of the VKG naive implementation and to reduce the end-to-end latency and enhance the overall performance of our chatbot. Indeed, an end-to-end latency of $\sim$7s, as the one attainable by the best configuration in SoA \cite{khan2025exasage} (see Table~\ref{tab:chatbot_implementations}), will make the DA chatbot end-user experience negative. Table~\ref{tab:latency_vkg_versions} reports the average latencies of the VKG-naive and VKG-optimized implementations, with a breakdown by sub-task (see Section~\ref{sec:VKG}) for both IoT datalake configurations: NoSQL DB and Parquet files.

In the VKG-naive implementation with the NoSQL DB as the IoT datalake, the average total VKG latency is 8.90s, with sub-task latencies as follows: Entity Extraction – 0.01ms, Data Fetching – 3.73s, VKG Creation – 3.92s, and GraphDB Storing – 1.25s. For the Parquet-based IoT datalake, the average total VKG latency is 6.78s, with Entity Extraction – 0.01ms, Data Fetching – 1.67s, VKG Creation – 3.87s, and GraphDB Storing – 1.24s. 

Since the GraphDB Storing task simply uploads the generated VKG via the GraphDB API, it offers limited room for optimization. Likewise, the Entity Extraction task has negligible runtime. Therefore, our optimization efforts are focused on the Data Fetching and VKG Creation tasks. However, in the case of the NoSQL DB as IoT datalake, the Data Fetching task—executing templated NoSQL/SQLite queries and retrieving data from a remote source—is bounded by the database API and network latency, which limits the impact of the proposed optimizations in the overall performance.

The first VKG optimization proposed targets the Data Fetching task. In the VKG-Naive implementation, data handling and processing were performed using the Python Pandas library. While Pandas is well-suited for rapid prototyping, it is not the most efficient in terms of execution speed. To address this limitation in the Data Fetching task, we adopt the Polars library. Built on a Rust backend, Polars offers significant performance advantages through multi-threading, columnar storage, and lazy evaluation—enabling substantial improvements in both speed and energy efficiency \cite{jetbrains2024polarsvspandas, polars2024benchmarkenergy}. This reduces the average time for the Data Fetching task by 92.75\% for the parquet IoT datalake, compared to the VKG-naive implementation.

\begin{table}[h]
\centering
\caption{Benchmarking time and storage size of RDF serialization formats for VKG creation with 5 million RDF triples.}
\label{tab:benchmark_serialization}
 \resizebox{0.8\columnwidth}{!}{%
\begin{tabular}{|l|r|r|}
\hline
\textbf{Serialization Format} & \textbf{Time [s]} & \textbf{Storage Size [MiB]} \\ \hline
N-Triples (NT) & 20.55 & 384.57 \\ \hline
Turtle (TTL) & 235.25 & 234.49 \\ \hline
XML & 93.11 & 625.12 \\ \hline
\end{tabular}
}
\end{table}

The second optimization we propose targets the VKG Creation task, which is the most time-consuming task of the VKG-naive implementation across both IoT datalake configurations. During this task, the collected data—represented as RDF triples—is integrated into a VKG and then serialized, i.e., converted into a specific format (e.g., Turtle, XML, or N-Triples) for storage. To determine the most time-efficient serialization format, we benchmarked the serialization of 5 million randomly generated triples (approximately 16\% larger than the biggest VKG in our dataset; see Table~\ref{tab:perf_metrics}). As shown in Table~\ref{tab:benchmark_serialization}, N-Triples offers the fastest serialization time at 20.55s, producing a file of 384.57 MiB. In comparison, Turtle—while more compact and human-readable—requires 235.25s and yields a 234.49 MiB file, whereas XML takes 93.11s and generates a 625.12 MiB file.

Given that our primary objective is to reduce latency, execution speed takes precedence over file compactness. N-Triples provides the best trade-off, being 91.3\% faster than Turtle and 77.9\% faster than XML. While it produces files 64\% larger than Turtle, it remains 38.4\% smaller than XML. Furthermore, since GraphDB parses and stores uploaded data in its own optimized binary format, the choice of input serialization format does not affect final storage efficiency. Therefore, N-Triples is selected as the optimal serialization format for our VKG-optimized implementation. In addition, another improvement was made by collecting all generated triples in a list first, then adding them to the RDF graph at the end. In the VKG-naive, each triple was added to the graph immediately after its creation, which proved less efficient and slowed down the process.

Following these optimizations, the average total VKG latency for the VKG-optimized implementation with the NoSQL DB as the IoT datalake is 6.52s, with sub-task latencies as follows: Entity Extraction – 0.01ms, Data Fetching – 3.73s, VKG Creation – 1.54s, and GraphDB Storing – 1.25s. For the Parquet-based IoT datalake, the average total VKG latency is 2.90s, with Entity Extraction – 0.01ms, Data Fetching – 0.12s, VKG Creation – 1.54s, and GraphDB Storing – 1.24s. 

The VKG-optimized implementation with the NoSQL DB IoT datalake shows a 26.7\% improvement in total VKG latency compared to the VKG-naive with the NoSQL DB IoT datalake. Similarly, the VKG-optimized with Parquet files as the IoT datalake demonstrates a 57.2\% improvement in total VKG latency over the VKG-naive with the Parquet IoT datalake. These results highlight a significant improvement in the VKG generation process for both IoT datalake configurations.

The above analysis was conducted on average values. Table~\ref{tab:perf_metrics} provides detailed statistical measures, including the mean, standard deviation (std), minimum (min), first quartile (25\%), median (50\%), third quartile (75\%), and maximum (max)  of the VKG-optimized implementation for the following: number of triples generated, number of nodes, data points fetched, storage size, and execution times of individual tasks. We focus the analysis on the Parquet IoT datalake configuration as it is the most performing one. Notable observations include a mean of 90,512 triples in the generated VKG with a maximum of 4,299,260, and an average of 30,666 nodes with a peak of 862,423. The average number of data points fetched is 61,371, with a maximum of 2,575,440. Regarding storage, the mean storage size is 3.71 MiB, with the largest entry reaching 179.4 MiB. Execution time is broken down into the four VKG sub-tasks and the overall end-to-end VKG latency. The Entities Extraction task is nearly instantaneous, with negligible variation and execution times consistently under 0.01ms. Data Fetching task averages 0.12s, but can take up to 1.86s in more complex cases. VKG Creation task and GraphDB Storing task show greater variability, with mean durations of 1.54s and 1.24s, respectively, and occasional spikes up to 49.31s and 92.48s. Total VKG latency reflects this variability, averaging 2.89s and reaching a maximum of 156.91s. 

\begin{table*}[t]
\centering
\caption{Summary statistics for key metrics—triples, nodes, data points, storage size, and execution times—showing mean, std, min, 25\%, 50\%, 75\%, and max for VKG-optimized with Parquet as the IoT datalake.}
\label{tab:perf_metrics}
 \resizebox{\textwidth}{!}{%
\begin{tabular}{|l|r|r|r|r|r|r|r|}
\hline
\textbf{Metric} & \textbf{Mean} & \textbf{Std} & \textbf{Min} & \textbf{25\%} & \textbf{50\%} & \textbf{75\%} & \textbf{Max} \\ \hline
\# Triples in   VKG & 90,512 & 476,365 & 5 & 701 & 2,572 & 12,842 & 4,299,260 \\ \hline
\# Nodes in   VKG & 30,666 & 132,991 & 6 & 372 & 1,303 & 5,755 & 862,423 \\ \hline
Data Points   Fetched & 61,371 & 303,182 & 4 & 364 & 1,503 & 8,416 & 2,575,440 \\ \hline
Storage size   [MiB] & 3.68467 & 19.66756 & 0.00029 & 0.02621 & 0.09558 & 0.54271 & 179.40454 \\ \hline
Time: Entities   Extraction [s] & 0.00001 & 0.00001 & 0.00001 & 0.00001 & 0.00001 & 0.00001 & 0.00001 \\ \hline
Time: Data   Fetching [s] & 0.12104 & 0.25397 & 0.00001 & 0.05687 & 0.09603 & 0.11891 & 1.86240 \\ \hline
Time: VKG   Creation [s] & 1.53804 & 5.90471 & 0.47102 & 0.47631 & 0.48924 & 0.57034 & 49.30640 \\ \hline
Time: GraphDB   Storing [s] & 1.24045 & 9.31868 & 0.00767 & 0.01684 & 0.03225 & 0.13819 & 92.48742 \\ \hline
Time: Total VKG Latency [s] & 2.89954 & 16.29448 & 0.01391 & 0.03263 & 0.12714 & 0.43571 & 156.91280 \\ \hline
\end{tabular}%
}
\end{table*}

\subsection{Single user question response time}

\begin{figure}[h!]
    \centering
    \includegraphics[width=\columnwidth]{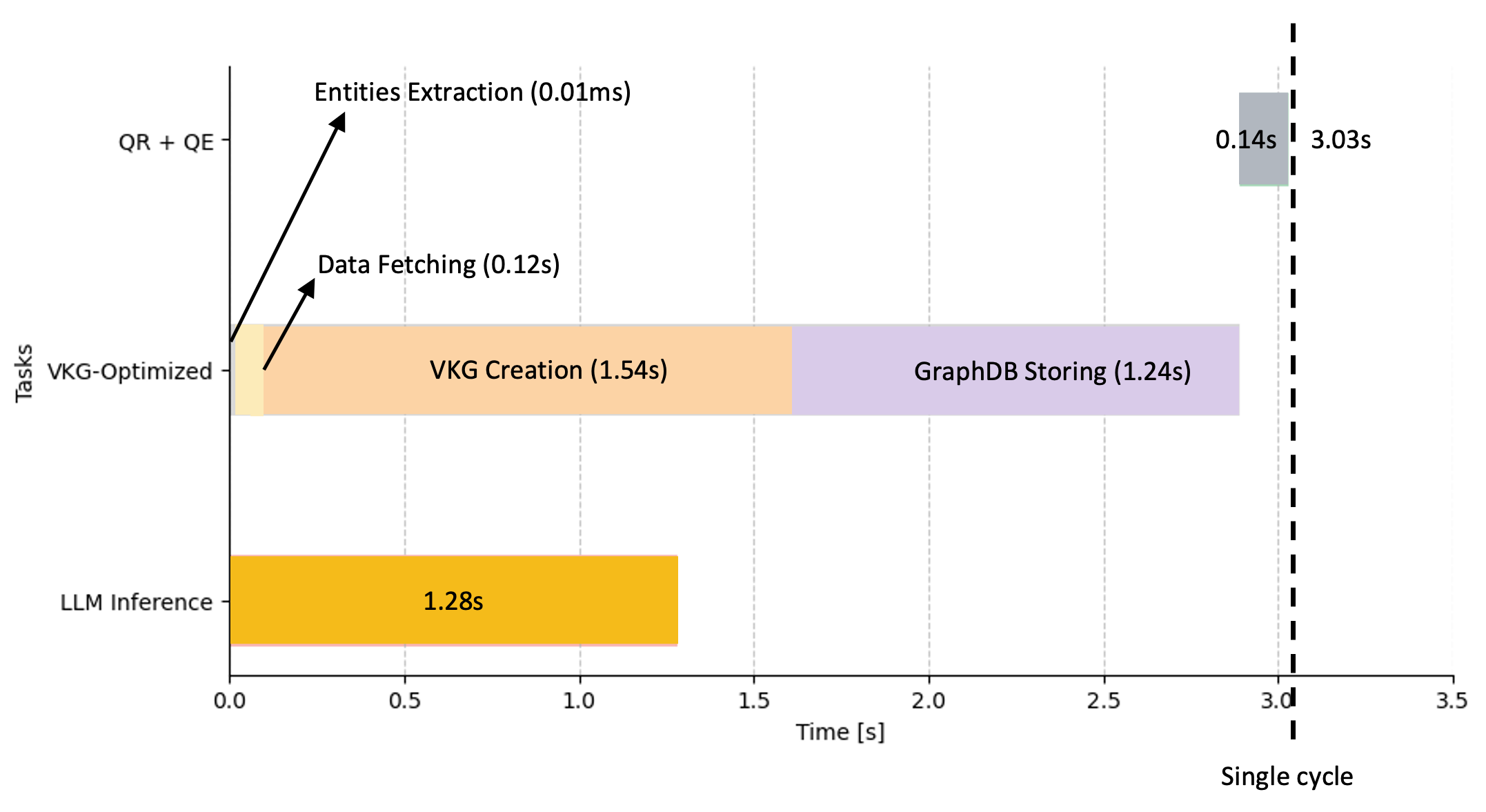}
    \caption{Timeline of a single user question cycle showing average latencies for QR, QE, LLM Inference, and VKG generation (with sub-tasks) in the proposed DA Chatbot with a Parquet-based IoT datalake.}
    \label{fig:gantt_chart}
\end{figure}

Figure~\ref{fig:gantt_chart} illustrates the timeline of a single user question execution. Upon receiving a question from the user, the backend triggers LLM Inference and VKG generation concurrently. On average, LLM Inference takes 1.28s, and VKG generation takes 2.89s. Once both processes are complete, the backend proceeds with query refinement and execution (QR + QE in the figure), which takes an additional 0.14s. The total time for processing a single user question on average is thus 3.03s.

\section{Current Limitations and Future Directions}
\label{sec:current_limitations}

One major drawback of the Algorithm~\ref{alg:process-prompt} is the algorithm's dependence on predefined patterns, which limits its ability to generalize across diverse and abstract user questions. For instance, when users ask questions using conceptual terms—such as “average power consumption”—the algorithm may fail to map the word “power” to the correct metric if that metric is not explicitly listed in the plugin-metric mapping derived from the sensor metadata. This restricts the algorithm's understanding to only those terms explicitly encoded in the system, thus failing to capture semantically similar or higher-level descriptions. 

Temporal expressions present a similar challenge. Natural language allows for a wide variety of temporal phrasing (e.g., “last 10 hours”, “past week”, or “yesterday afternoon”), which may not conform to the strict [YYYY-MM-DD HH:MM:SS] format expected by the algorithm. As a result, the system often fails to correctly interpret or extract time-related information unless the input precisely matches the known regex patterns.

To partially address this limitation, one potential improvement involves adding interactivity to the chat interface. If the algorithm fails to match a known entity (e.g., a sensor name or metric), it could proactively suggest likely candidates and prompt the user for confirmation. For example, if a user requests a “power metric”, the system could retrieve all relevant metrics associated with power consumption and present them for user selection. This interactive feedback loop could also apply to other entity types, thereby enhancing system robustness. However, while this interactive design improves accuracy, it may reduce the overall user experience by placing more cognitive load on the user—something we aim to avoid in the development of the DA Chatbot. 

A more scalable and user-friendly solution involves the integration of language models for entity extraction. LLMs, such as GPT-4o, inherently understand abstract concepts like “power” and temporal expressions, offering a more context-aware and semantically rich approach to parsing user questions. Preliminary experiments using OpenAI ChatGPT models to perform the same entity extraction task as Algorithm~\ref{alg:process-prompt} have shown promising results. However, a comprehensive evaluation is necessary to fully validate their performance, reliability, and end-to-end latency.

\begin{figure}[h]
    \centering
    \includegraphics[width=0.7\columnwidth,trim=0 0 0 0.7cm,clip]{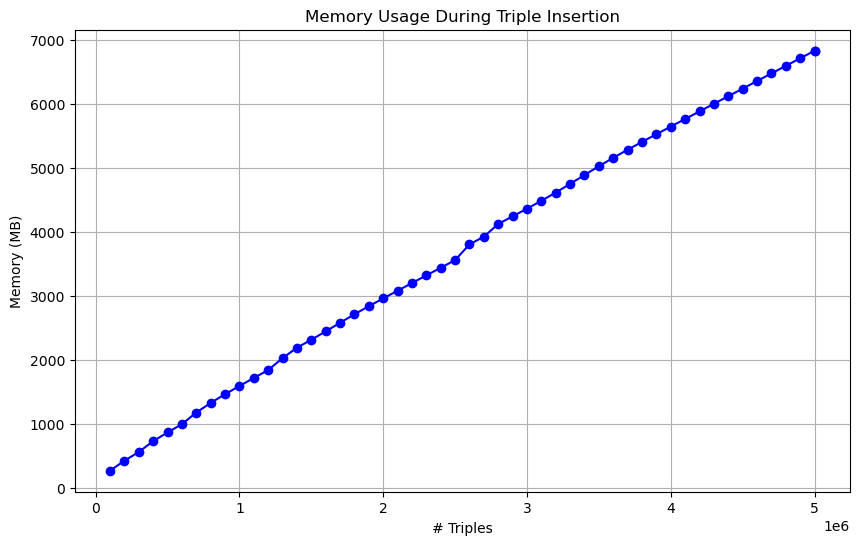}
    \caption{Memory usage vs. number of RDF triples added to an in-memory RDF-graph, evaluated up to 5 million triples.}
    \label{fig:mem_benchmark}
\end{figure}

In addition to limitations in natural language understanding, the system faces significant scalability constraints in the VKG Creation task: the creation of the VKG through RDF triple generation. In our implementation using Python’s rdflib for RDF graphs, all triples are generated and stored in memory, leading to increasingly high memory consumption as the graph grows. As shown in Figure~\ref{fig:mem_benchmark}, memory usage exhibits a consistently increasing trend with the number of triples added. For 5 million triples, the memory utilization reaches approximately 6.36 GiB.

To address this, future versions of the VKG generation pipeline should include adaptive memory management strategies. One such approach is to introduce chunking, where triples are generated and written to temporary storage in manageable batches, rather than kept entirely in memory. These chunks could then be sequentially uploaded to the triplestore (GraphDB Storing task), thereby reducing peak memory usage and enhancing system stability.

GraphDB Storing task—uploading RDF triples to the triplestore—is itself a bottleneck when the volume of triples is large. While commercial versions of GraphDB support concurrent uploads, the free version used in our current setup does not. This limits throughput and increases processing time. To address this limitation, we propose two potential solutions: (i) exploring alternative open-source triplestores, such as Apache Jena, which support parallel ingestion of RDF data and offer improved scalability for large-scale deployments; or (ii) migrating to more modern property graph and more performant databases like Neo4j.

Collectively, these enhancements—in natural language understanding, memory efficiency, and data upload optimization—outline a clear roadmap for advancing the proposed DA chatbot and its underlying VKG generation architecture.

\section{Conclusion}
\label{sec:conclusion}

We present the first end-to-end Data Analytics (DA) chatbot for IoT telemetry, capable of translating natural language queries into SPARQL over dynamically constructed Virtual Knowledge Graphs. Using data center IoT telemetry as a case study, we demonstrate its effectiveness. This KG-based approach achieves 92.5\% accuracy, far surpassing LLM-to-NoSQL methods (25\%), while avoiding the storage cost of a full static KG. Our architecture integrates a frontend, a Flask API-based backend, an LLM inference service, an IoT datalake, and a graph database, delivering a 3.03\,s average end-to-end latency—an 85\% improvement over prior work. This design enables intuitive, scalable, and real-time access to large-scale IoT telemetry. Future work will address natural language generalization, temporal reasoning, and large-scale VKG generation to further improve robustness and responsiveness.

\section*{Acknowledgment}

This research was partly supported by EuroHPC JU SEANERGYS (g.a. 101177590), HE EU DECICE project (g.a. 101092582), HE EU Graph-Massivizer project (g.a. 101093202), and Spoke “FutureHPC \& BigData” of the ICSC-Centro Nazionale di Ricerca in “High Performance Computing, Big Data \& Quantum Computing”, funded by the EU – NextGenerationEU. We also thank CINECA for their collaboration and for providing access to their computing resources.




\bibliographystyle{IEEEtran}
\bibliography{IEEEabrv,references}

\end{document}